# Optical dielectric Huygens' metagrating performing near-unity anomalous refraction at TM mode with extremely simple design

Rui (Ray) Yao*

Department of Electrical Engineering

City University of Hong Kong, Hong Kong, China

Email address: *ruiyao7-c@my.cityu.edu.hk

## Abstract

We numerically demonstrate a highly efficient optical Huygens' metagrating with unprecedentedly simple structure (only one meta-atom per period) designed via aggressively discretized method which initially originated from discretized metasurface design, performing nearly lossless anomalous refraction under TM-polarized incident light. A 2D full-wave Floquet simulation shows that the proposed metagrating anomalously transmits an incident light of 800 nm wavelength with up to 94% power efficiency, where the specular transmission is substantially suppressed. Our findings can also help demarcate the boundary between metagrating and metasurface.

Metasurfaces are two dimensional thin planar nanostructures which could serve as optical components with intriguing functionalities and potentials in manipulating the behavior of light waves, overwhelming their conventional bulky counterparts whose light manipulation capability are heavily reliant on the accumulative effect along the thicker optical path [1-6]. In recent years, loads of efforts have been invested in the field of metasurface aimed at fully unleashing its potentials as a new paradigm for controlling the electromagnetic (EM) waves in a nearly arbitrary manner [7], leading to a bundle of applications related, including meta-hologram [8], meta-lens [9, 10], LED light extractor [11], flat-optics [4], differential equation solver [12], diffraction mode circulator [13], anomalous reflection/refraction [14-17], beam splitter [5], among many others.

Although having the success of achieving many fancy applications, the functional mechanism behind its novel ability of tailoring the behavior of EM wave is primarily based on the phase gradient approach originated from the generalized snell's law proposed by N. Yu et al. [6], based on which we can artificially engineer the reflection/transmission coefficient and phase, redirect the incident wave into desired directions by imparting the required tangenti al momentum to it [6]. However, due to the rapid development in this field, there have been many challenges facing this seemingly perfect theory with researchers pointing out the intrinsic limitation on this approach which often results in low efficiency and highly compact layout of the meta-atoms (constituent elements in the metasurface) severely exacerbating the complexity in fabrication. For instance, the main problem of wavefront steering happens when designing metasurfaces for extreme deflection angles, in which this approach is inevitably afflicted by a compromise made between efficiency and steering angles. Because this design approach only focuses on geometric phase difference, lacking the consideration of impedance mismatch between incident wave and the wavefront redirected [18].

In a bid to address the aforementioned deficiencies, Huygens' metasurface – performing wavefront transformation with the help of high permittivity dielectric meta-atoms and overlapped electric/magnetic dipole mode within the resonators, have been proposed, exhibiting the nature of reaching full transparency to the incident wave and complete $2\pi$ phase coverage [5, 19]. Following the footstep, which ultimately led to the birth of aggressively discretized metasurface consisting of smaller number of relatively larger unit cells per period and being able to perform extreme wavefront manipulation with high efficiency and fabricating simplicity [2, 3, 13, 14]. The larger unit cell can substantially alleviate the mutual coupling effect between the adjacent meta-atoms, which otherwise be strong in its densely packed counterparts. The notion of aggressively discretized metasurface stems from long-existing classical grating physics, where the metasurface with periodic structure can generate a discrete series of propagating and evanescent plane waves, known as Floquet-Bloch (FB) modes, allowing us to eliminate the spurious FB modes while retaining the desired FB modes excited by the incident wave through judiciously designing its period, meta-atoms and unit cell size. The metasurface designed in this manner can see its desired propagating mode being strongly enhanced and the FB mode operation in k-space domain.

In this paper, we surprisingly found that the aforementioned metasurface design approach can be further simplified, making it resemble a newly arrived concept: "metagrating", a grating-like meta-device with complex single grating elements performing similar wavefront manipulation as the ordinary metasurfaces. Metagratings have been hailed as the simplest meta-device, an ideal upgrade from metasurfaces with negligible mutual coupling effect, enabling near unity efficiency at wavefront manipulation without the need of deeply subwavelength constituent elements, ending the reliance on continuous phase change or surface impedance matching [15, 16, 20]. Here, we numerically introduce a dielectric Huygens' metagrating designed by exploiting the aggressively discretized method but comprising only one silicon meta-atom in cylindrical shape per period, which performs anomalous refraction with nearly perfect efficiency by redirecting an TM-polarized incident light wave of 800 nm wavelength at $\theta_i = 50°$, 30° into an anomalously refracted light wave at $\theta_t = -22.5°, -18.7°$. The example of the metagrating design presented in this paper is systematic, being able to be upscaled to longer wavelength regime such as far-infrared and terahertz. To the best of our knowledge, the proposed dielectric Huygens' metagrating is the first of this kind with the simplest ever structure featuring only one meta-atom per period and achieves the maximum refraction coefficient up to 0.97.

Fig. 1 shows a schematic of the overall dielectric Huygens' metagrating (DHMG) and the geometry of one period of the DHMG, in which the metagrating consists of an 2D array of silicon dielectric meta-atom in cylindrical shape. Each meta-atom shown in Fig 1(b) acts as Huygens source based on schelkunoff's equivalence principle, in which the mutually orthogonal electric/magnetic dipole mode can be overlapped by finely tuning its radius and height, prohibiting any reflection from the surface and allowing all incoming waves to pass through the surface [19]. Note that the dielectric meta-atom uniquely enables the meta-grating/surface to perform transmitting functionality with single layer structure compared to its metallic counterparts where multi-layer structure is inevitable [1, 13].

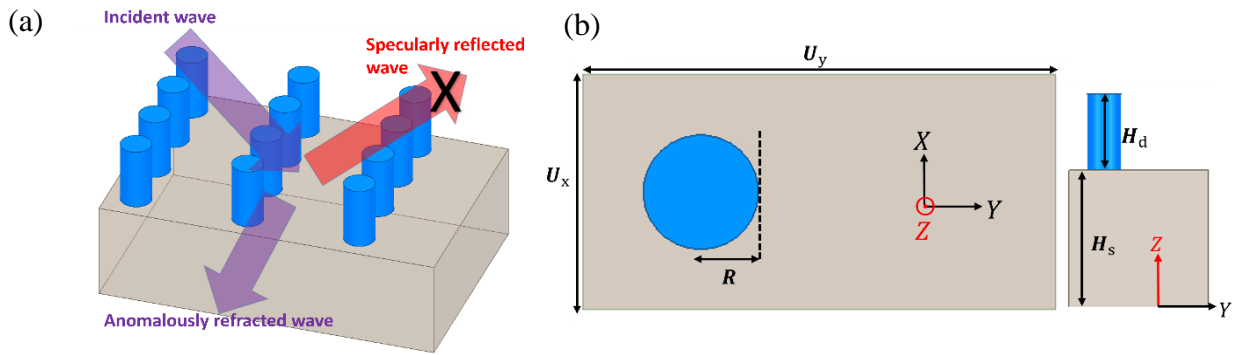

**Fig. 1.** (a) Overview of the dielectric Huygens' metagrating formed by 2D array of cylindrical dielectric meta-atoms. (b) The top-view and side-view of one period (supercell) of the proposed metagrating. The optimized supercell dimensions are: $U_y = 0.87\lambda_0$, $U_x = 0.435\lambda_0$, $R$ (radius of the meta-atom) $= 90\ nm$, $H_d = 385\ nm$, $H_s =$ sufficiently thick.

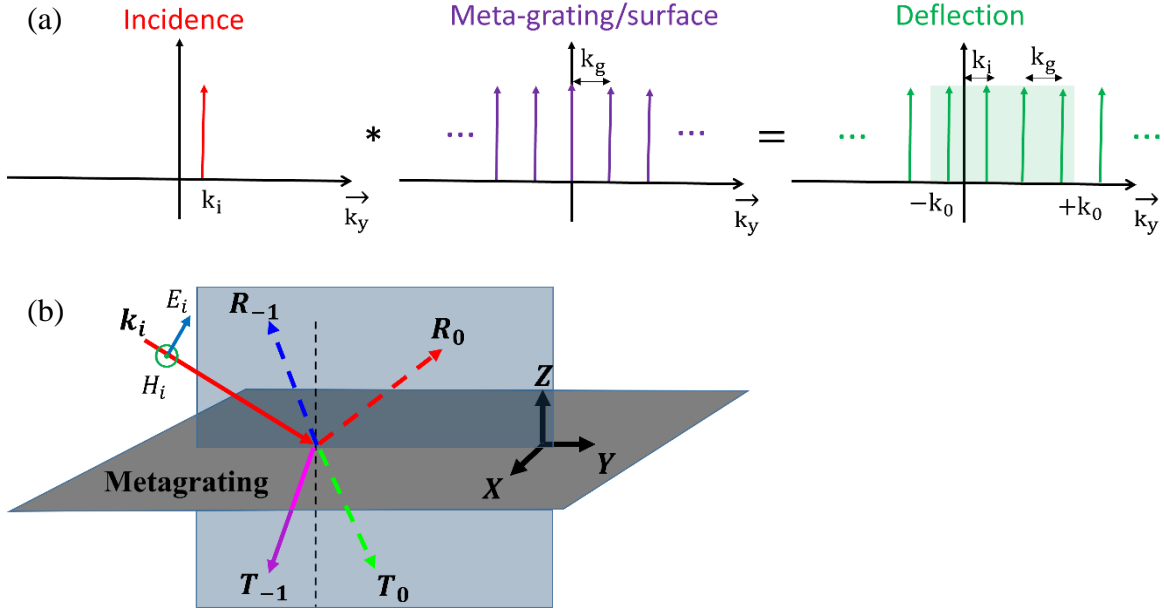

**Fig. 2.** (a) The $k$-space operation of the periodic meta-grating/surface generating diffraction modes. The arrows (irrelevant to the amplitude) represent the existing FB modes, the greenish box contains the allowed reflected/transmitted modes with the evanescent modes located outside. (b) A schematic of the designed metagrating with all FB modes: incident wave ($k_i$), desired propagating FB mode ($T_{-1}$), specular transmission ($T_0$), specular reflection ($R_0$) and first order reflection ($R_{-1}$). The solid red arrow indicates a TM mode incident wave with electric field oscillating in y-z plane.

Next, we proceed to design the metagrating operating at $\lambda_0 = 800$ nm wavelength using the aforementioned "aggressively discretized approach" as if we are designing a metasurface. The "aggressively discretized approach" can be understood by considering a periodic meta-surface/grating in free-space with period $\Lambda_g$ and spatial frequency $k_g = \frac{2\pi}{\Lambda_g}$. Under the illumination of a plane wave, a discrete set of propagating and evanescent Floquet-Bloch (FB) modes are excited along the meta-surface/grating. The transverse spatial frequency of these FB modes can be expressed as:

$$k_{y(m)} = k_i + mk_g = k_i + m\frac{2\pi}{\Lambda_g}, \quad (1)$$

Where $k_i$ indicates transverse wavenumber of incident wave projected in y-axis and $m = 0, \pm 1, \pm 2, \cdots$ (an integer) represents the corresponding FB mode. $\Lambda_g, k_g$ are the period of the meta-surface/grating and the spatial frequency. Fig. 2(a) shows the operation in spectrum domain (or K-space) for a meta-surface/grating with the mathematical expression as:

$$\Omega_o(k_y) = \sum_m A_m \delta(k_y - k_{y(m)}), \qquad (2)$$

Where $\Omega_o(k_y)$ is the K-space spectrum of the wave transmitted, $A_m$ is the complex amplitude, $k_{y(m)}$ is given in Eq. (1). As shown by the arrows in Fig. 2(a), the spectrum of the transmitted wave has an infinite amount of FB modes, but only a finite number amount located within the propagating region, in which $k_y \in [-k_0, k_0]$ (enclosed by a greenish box), can be scattered into far-field, whereas the modes outside the propagating region are evanescent waves which are confined in near-field. It has been proved that the aggressively discretized metasurface with M-meta-atoms per period suffices to redirect the reflected/transmitted power in the $m_{th}$ - FB modes into the far-field [14]. Such metasurface can achieve anomalous reflection/transmission with spurious and specular modes being substantially suppressed via judicious design. The PB-modes lying within the aforementioned propagating region that can reradiate into the $m_{th}$ physically allowed directions at far-field is expressed as:

$$\sin \theta_{t(m)} = \sin \theta_i + \frac{m \lambda_0}{\Lambda_g}, \qquad (3)$$

Where $\theta_i$ is the incident angle, $m$ is again the FB mode number and $\Lambda_g$ is the periodicity of the metasurface. We seek to construct an aggressively discretized meta-surface/grating performing anomalous refraction that can efficiently redirect the power carried by the incident wave into the allowed $m = -1$ FB mode with the transmitted angle of $\theta_{t(-1)}$ while assuring that $m = -1$ is the only existing high order mode in addition to the fundamental mode ($m = $ 0, specular transmission in this case). In order to pinpoint the level of discretization, number of meta-atoms per period and the metasurface's periodicity. We resort to the relation[14]:

$$N = 1 + \left[\frac{(1 - \sin\theta_i)}{|\sin\theta_t - \sin\theta_i|}\right] + \left[\frac{(1 + \sin\theta_i)}{|\sin\theta_t - \sin\theta_i|}\right] \quad (4)$$

Where N indicates the number of propagating FB modes as well as number of meta-atoms per period of the metasurface. In this case, we take $N = 2$ to acquire the minimum number of meta-atoms per period and redirect most of the incident power into $m = -1$ PB mode while maximally suppressing the direct transmission and both first order and specular reflection, although our finalized metagrating consists of only one meta-atom per period without compromising any desired functionalities mentioned above.

As a design example, we design our refractive dielectric Huygens metagrating operating at 800 nm wavelength which receives a TM-polarized incident light at $\theta_i = 50°, 30°$ and diverts them into the first order FB mode ($m = -1$) with refracted angles of $\theta_t = -22.5°, -18.7°$ calculated under the constraint of $N = 2$ from Eq. (4). Afterwards, substituting these angles into Eq. (3) arrives at the metasurface's periodicity in y-direction as $\Lambda_g = \frac{\lambda_0}{|\sin\theta_i - \sin\theta_{t(-1)}|} = 0.87\lambda_0$. Fig 2.(b) shows the operation of the dielectric metagrating performing anomalous refraction with 4 available modes: desired propagating FB mode (T$_{-1}$), specular transmission (T$_0$), specular reflection (R$_0$) and first order reflection (R$_{-1}$) highlighted. Pragmatically, the designed mete-surface/grating is reflectionless attributed to the intrinsic property of dielectric Huygens' metasurface such that the reflection related modes are ideally eliminated [19], nevertheless there modes might be weakly excited. The simulation is carried out using full wave Ansys HFSS simulator with the silicon meta-atom [Fig 1.(b)] in cylindrical shape placed in an infinite air box (unit cell) in which the x and y directions are assigned with periodic boundary condition. The topmost and bottom surfaces are imposed by Floquet port generating an EM plane wave propagating along $-z$ direction, which illuminates the meta-atom at $\theta_i =$ 50°, 30°. The frequency dependent complex refractive index of silicon with losstangent counted is taken from [21]. A sufficiently thick SiO₂ substrate built below the meta-atom to emulate the real scenario with a fixed refractive index of 1.4 [5]. By default, we tune the radius and height of the meta-atom to create a collection of meta-atoms having different transmission phase and similar transmission magnitude, then we combine two meta-atoms chosen from the collection with high transmission and a phase difference of 180° to form one period of the aggressively discretized metasurface (ADM), fine-tuning of the two meta-atoms is needed to counter their mutual coupling effect.

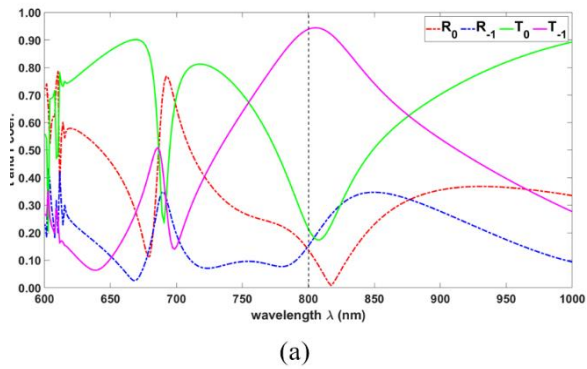

(a)

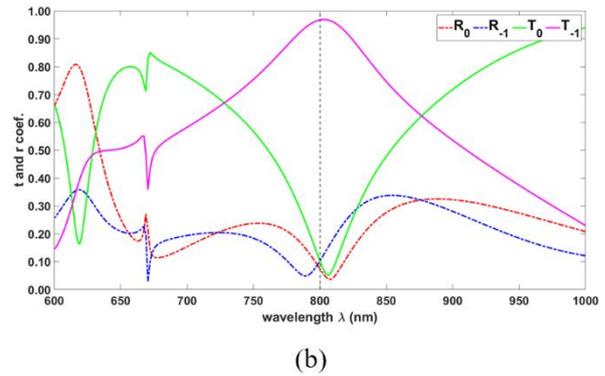

(b)

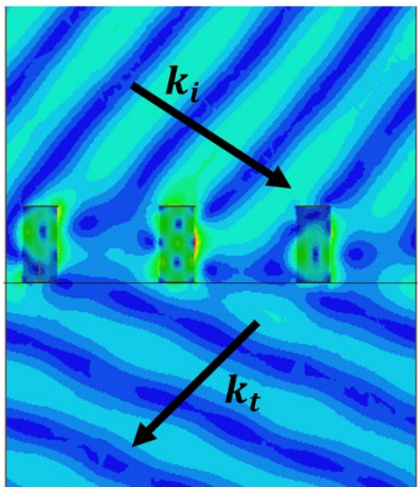

(c)

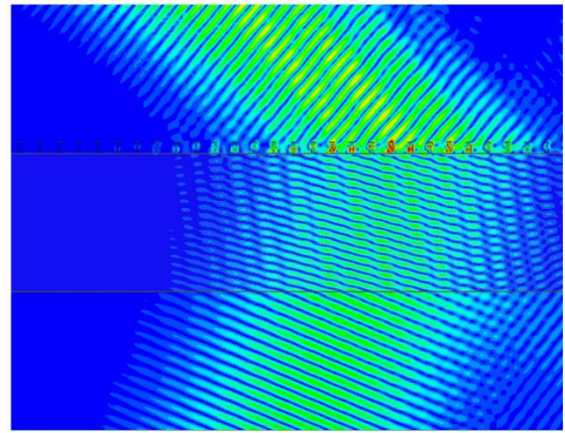

(d)

**Fig. 3.** (a), (b) Spectra of transmission/reflection coefficient of the -1 and $0_{th}$ FB modes illuminated by incident waves of $\theta_i = 50°, 30°$, respectively, with the black dashed line intercepting the mode curves at the operating wavelength (800 nm), where the highest efficiency occurs at $\theta_i = 30°$. (c) Simulated electric field distribution in y-z plane behaving anomalous refraction with 3 supercells employed. (d) Electric field distribution produced by "finite fa-field simulation" excited by a gaussian beam with 200 supercells repeatedly placed in y-direction. The upper region shows the incident wave propagating in the air, impinges the metagrating at $\theta_i = 50°$ with minimal reflection, the anomalously refracted wave propagates within the $SiO_2$ substrate (middle region) and the outgoing wave going into the air (bottom region).

Inspired by the concept of metagrating, we unexpectedly uncovered the compatibility of using standardized ADM approach in metagrating design by removing one meta-atom from each period and judicious optimization, resulting in an extremely concise supercell with only one meta-atom per period as shown in Fig 1.(b). Note that the designed metagrating achieving exceptionally high efficiency outperforms its metasurface counterpart under the illumination of a TM-polarized incident wave with electric field oscillating in y-z plane [Fig 2(b)]. The transmission and reflection coefficient corresponding to the propagating FB modes are presented in Fig 3(a), (b) respectively, explicitly showing that the maximum anomalous refraction efficiency occurs at 800 nm wavelength with nearly 94% of the power being coupled into the desired FB mode ($T_{-1}$) and the specular transmission ($T_0$) mode substantially suppressed. Furthermore, the designed dielectric metagrating is nearly transparent with less than 2% of the power contributing to reflection mode ($R_0$, $R_{-1}$). The simulated electric field distribution displayed in Fig 3. portrays the behaviour of the desired anomalous refraction. Especially, our self-developed "finite far-field simulation" [Fig 3.(b)] employing an gaussian beam to excite the model with finite length better imitates the practical working environment.

The consistent result of these two simulations for electric field distribution further endorses the functionality of our designed metagrating.

In conclusion, we numerically devised a dielectric Huygens' metagrating operating at NIR regime, capable of anomalously rerouting the TM-polarized light wave incoming at $\theta_i = 50°$, 30° into the refracted light wave at $\theta_t = -22.5°, -18.7°$ with nearly unity efficiency, based on a systematic design approach originally bound to designing aggressively discretized metasurface. To date, it is the first of this kind combining the simplest ever structure and the highest efficiency into one, which enables it to be up/down-scaled to millimeter-wave, terahertz-wave or infrared, visible frequency domain. The unique design approach implemented herein opens up a brand-new realm in developing novel meta-devices and facilitates the potential of photonic integrated circuit system, more importantly it might provide us with an insight on better marking the demarcation between metagrating and metasurface.